\documentclass[transmag,12pt]{IEEEtran}
\usepackage[shortlabels]{enumitem}
\usepackage{graphicx} 
\usepackage{epsfig} 
\usepackage{amsmath,amsthm,amssymb}  
\usepackage{float}
\usepackage{cases,setspace,adjustbox}
\usepackage{cite}
\usepackage{array}
\usepackage[update,prepend]{epstopdf}
\usepackage{eqparbox}
\usepackage{url}
\usepackage{caption, subcaption}
\usepackage{xcolor}
\usepackage{mathtools}
\usepackage{bbm}
\usepackage{multirow}
\usepackage{multicol}
\usepackage{hyperref}

\usepackage{booktabs}

\usepackage[acronym]{glossaries}
\usepackage{comment}
\setacronymstyle{long-short}

\newacronym{AR}{AR}{Augmented Reality}

\newacronym{cdf}{CDF}{cumulative distribution function}
\newacronym{ctl}{CTL}{Communications Technology Laboratory}
\newacronym{cp}{CP}{Cyclic Prefix}
\newacronym{comp}{CoMP}{Coordinated Multi-Point}

\newacronym{d2d}{D2D}{Device-to-Device}
\newacronym{dl}{DL}{downlink}
\newacronym{dc}{DC}{dual connectivity}
\newacronym{dsrc}{DSRC}{Dedicated short-range communications}

\newacronym{embb}{eMBB}{enhanced mobile broadband}
\newacronym{enb}{eNB}{Evolved Node-B}
\newacronym{epc}{EPC}{Evolved Packet Core}
\newacronym{emtc}{eMTC}{enhanced MTC}

\newacronym{fdd}{FDD}{Frequency Domain Duplex}
\newacronym{fdm}{FDM}{Frequency Domain Multiplexing}
\newacronym{firstnet}{FirstNet}{First Responder Network Authority}

\newacronym{gnb}{gNB}{Next Generation Node-B}
\newacronym{gps}{GPS}{Global Positioning System}

\newacronym{harq}{HARQ}{Hybrid Automatic Repeat reQuest}

\newacronym{iiot}{IIoT}{Industrial Internet of Things}
\newacronym{its}{ITS}{Intelligent Transport Systems}
\newacronym{itu}{ITU}{International Telecom Union}

\newacronym{lte}{LTE}{Long Term Evolution}
\newacronym{lteapro}{LTE-A~Pro}{LTE Advanced Pro}

\newacronym{mcs}{MCS}{Modulation and Coding Scheme}
\newacronym{mcptt}{MCPTT}{Mission-Critical Push-to-Talk}
\newacronym{mimo}{MIMO}{Multiple-Input Multiple-Output}
\newacronym{mmtc}{mMTC}{massive machine-type communication}
\newacronym{mtc}{MTC}{Machine-Type Communications}

\newacronym{nist}{NIST}{National Institute of Standards and Technology}
\newacronym{nbiot}{NB-IoT}{Narrowband-Internet of Things}
\newacronym{nr}{NR}{New Radio}
\newacronym{ntia}{NTIA}{National Telecommunications and Information Administration}

\newacronym{psc}{PSC}{Public Safety Communications}
\newacronym{pscch}{PSCCH}{Physical Sidelink Control Channel}
\newacronym{psfch}{PSFCH}{Physical Sidelink Feedback Channel}
\newacronym{pssch}{PSSCH}{Physical Sidelink Shared Channel}

\newacronym{qos}{QoS}{Quality of Service}

\newacronym{ran}{RAN}{Radio Access Network}
\newacronym{rat}{RAT}{Radio Access Technology}
\newacronym{re}{RE}{Resource Element}

\newacronym{sl}{SL}{sidelink}

\newacronym{tdd}{TDD}{Time Domain Duplex}
\newacronym{tdm}{TDM}{Time Domain Multiplexing}

\newacronym{urllc}{URLLC}{ultra-reliable low-latency communication}
\newacronym{uav}{UAVs}{Unmanned Aerial Vehicles}
\newacronym{ue}{UE}{User Equipment}
\newacronym{ul}{UL}{uplink}

\newacronym{v2v}{V2V}{Vehicle-to-Vehicle}
\newacronym{v2x}{V2X}{Vehicle-to-Everything}
\newacronym{VR}{VR}{Vitual Reality}

\newacronym{wnd}{WND}{Wireless Networks Division}
\newacronym{wlan}{WLAN}{Wireless Local Area Network}
\newacronym{3gpp}{3GPP}{3rd Generation Partnership Project}
\newcommand{\ranA}{\text{RAN}_\text{A}}
\newcommand{\ranB}{\text{RAN}_\text{B}}

\begin{document}
\title{5G NR-LTE Coexistence: \\Opportunities, Challenges, and Solutions}
\author{Sneihil Gopal$^{\dagger\ddagger}$, David Griffith$^{\ddagger}$, Richard A. Rouil$^{\ddagger}$ and Chunmei Liu$^{\ddagger}$\\
$^{\dagger}$Department of Physics, Georgetown University, USA\\
$^{\ddagger}$National Institute of Standards and Technology (NIST), USA\\
Emails: \{sneihil.gopal, david.griffith, richard.rouil, chunmei.liu\}@nist.gov}
\maketitle
\section*{Abstract}
\label{sec:abstract}
5G New Radio (NR) promises to support diverse services such as enhanced mobile broadband (eMBB), ultra-reliable low-latency communication (URLLC), and massive machine-type communication (mMTC). This requires spectrum, most of which is occupied by 4G Long Term Evolution (LTE). Hence, network operators are expected to deploy 5G using the existing LTE infrastructure while migrating to NR. In addition, operators must support legacy LTE devices during the migration, so LTE and NR systems will coexist for the foreseeable future. In this article, we address LTE-NR coexistence starting with a review of both radio access technologies. We then describe the contributions by the 3rd Generation Partnership Project (3GPP) to solving the coexistence issue and catalog the major coexistence scenarios. Lastly, we introduce a novel spectrum sharing scheme that can be applied to the coexistence scenarios under study.\let\thefootnote\relax\footnotetext{U.S. Government work, not subject to U.S. Copyright.}

\section{Introduction}
\label{sec:intro}
The proliferation of multimedia applications and the rapid digitalization of industries has led to an exponential growth in mobile data traffic in the recent years~\cite{ref:ericsson_2022}. To cater to this increasing demand for data traffic, \gls{3gpp}, the global standards development organization, is currently specifying a fifth generation (5G) of radio interface referred to as \gls{nr}~\cite{ref:NR_Parkvall_2017}. 5G \gls{nr} will support increasing traffic demand and wireless connectivity for a wide range of new applications and use cases, such as automotive, healthcare, public safety, and smart cities. 
\begin{figure}[t]
    \centering
    \includegraphics[width=\columnwidth]{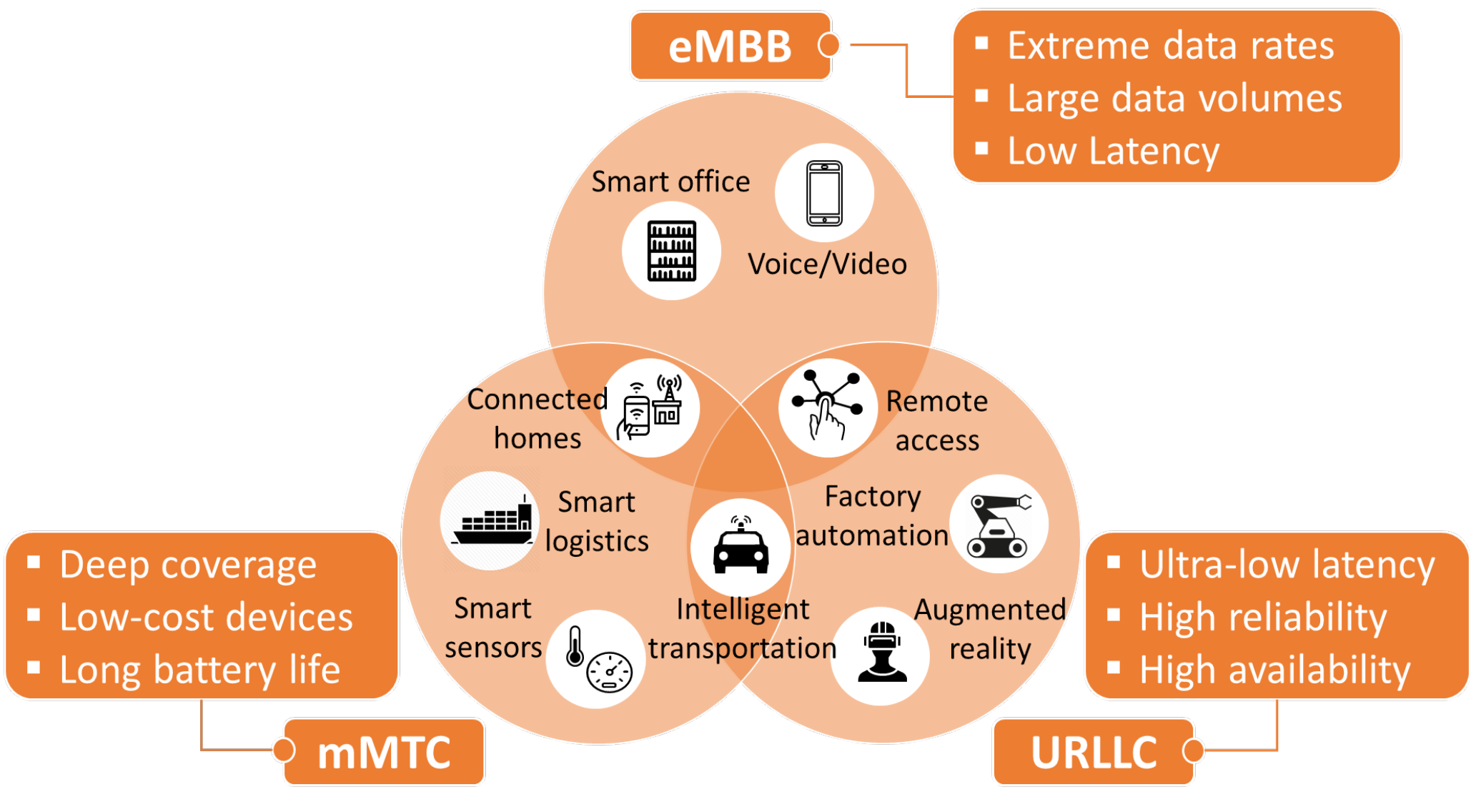}
    \caption{5G NR services, associated applications and their respective requirements.}
    \label{fig:services}
\end{figure}

Specifically, as illustrated in Figure~\ref{fig:services}, \gls{nr} will support three main use cases:~\gls{embb},~\gls{urllc}, and~\gls{mmtc}, each with different requirements and applications. \gls{embb} and~\gls{urllc} services require high data rates and high reliability with low latency, respectively, which in turn require large bandwidths such as what is available above 3~GHz. In addition,~\gls{embb} and~\gls{mmtc} services require good coverage to ensure network access for most users, which can be best achieved by operating below 2~GHz. To provide high data rate, low latency, and good coverage, 5G~\gls{nr} will operate in two spectral bands: Frequency Range 1 (FR1) from 410~MHz to 7.125~GHz, previously known as sub-6~GHz band, and Frequency Range 2 (FR2) from 24.25~GHz to 52.6~GHz, also known as mmWave band. 

Because almost all accessible low frequency bands are occupied by existing 4G \gls{lte}, operators have two choices: acquire new spectrum or refarm the existing spectrum from \gls{lte} to \gls{nr}. Both options are expensive and, since majority of the traffic in the near future will be carried by \gls{lte} networks, refarming low frequency bands from \gls{lte} without a corresponding increase in \gls{nr} devices will lead to congestion in the \gls{lte} bands and degrade network performance. Therefore, operators must leverage the existing 4G infrastructure as they migrate to \gls{nr}, while providing services to legacy devices. As a result, \gls{lte} and \gls{nr} networks will coexist for the foreseeable future.

In this article, we discuss the \gls{lte}-\gls{nr} coexistence issue with a focus on \gls{embb}, \gls{mmtc}, and \gls{urllc}, in the context of \gls{psc} networks, \gls{mtc} networks, and \gls{v2x} networks. We discuss the coexistence techniques being developed by \gls{3gpp}, and examine the major coexistence scenarios that can occur between \gls{lte} and \gls{nr} devices. Lastly, we propose a spectrum sharing scheme and discuss how a network operator could use the approach to support static and dynamic resource allocation for coexisting networks. 

In Section~\ref{sec:evolution}, we review the evolution of radio access technologies from \gls{lte} to \gls{nr}. In Section~\ref{sec:coexistence}, we address the \gls{lte}-\gls{nr} coexistence issue in different applications under study, describe the scenarios for each of them, and discuss \gls{3gpp}'s proposed solutions. Lastly, in Section~\ref{sec:scheme}, we introduce a novel and generic spectrum sharing scheme that is applicable to the coexistence scenarios under study.
\section{LTE to NR: An Overview}
\label{sec:evolution}
\begin{figure*}[t]
    \centering
    \includegraphics[width=\textwidth]{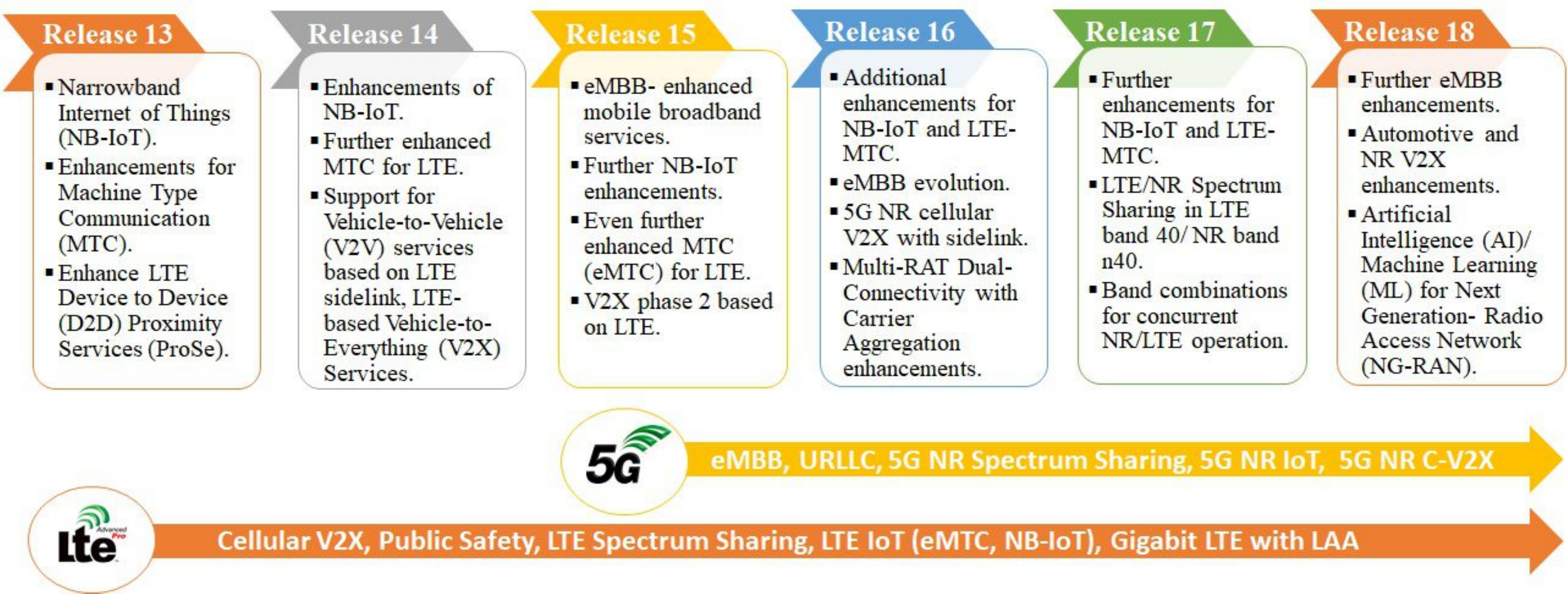}
    \caption{Evolution of radio access technologies from \gls{lte} to \gls{nr} and beyond.}
    \label{fig:Evolution}
\end{figure*}
\gls{3gpp} introduced \gls{lte} in Rel~8 with high spectral efficiency, variable bandwidths up to 20 MHz, and peak \gls{dl} and \gls{ul} data rates of 300 Mbit/s and 75 Mbit/s, respectively. Rel~9 introduced \gls{mimo} beam forming, multicast/broadcast services, and location-based services. 

In Rel~10, Rel~11, and Rel~12, also known as \gls{lte}-Advanced, \gls{3gpp} introduced several new features. In Rel~10, features such as carrier aggregation, \gls{ul} multiple antenna transmission, relaying, and enhancements to multicast/broadcast services were introduced, enabling peak \gls{dl} and \gls{ul} data rates of 3~Gbit/s and 1.5~Gbit/s, respectively. Rel~11 introduced \gls{comp} transmission/reception to improve coverage, cell-edge throughput, and spectral efficiency. Rel~12 included support for \gls{mtc} and public safety services such as \gls{d2d} communication, which is an enabling feature for \gls{v2x} communication.

Rel~13 and its successors are known as \gls{lte}-Advanced (\gls{lte}-A) Pro. Rel~13 introduced extended support for \gls{mtc} through \gls{nbiot} and enhanced \gls{mtc} (eMTC), enhancements to \gls{d2d} to support advanced proximity services for public safety services, and spectral efficiency enhancements via Full-Dimensional multiple-input multiple-output (FD-MIMO).  \gls{lte}-A Pro brought enhancements in multiple dimensions. It supported higher level of carrier aggregation and Licensed-Assisted Access (LAA), which led to the introduction of Gigabit \gls{lte} offering data rates up to 2~Gbit/s. It introduced enhancements to \gls{nbiot} and \gls{mtc}. It also expanded the reach of cellular technology to vehicular communication and introduced Cellular \gls{v2x} (C-\gls{v2x}), which is a crucial element of autonomous driving. The enhancements in Rel~15 made \gls{lte}-A Pro meet the International Telecom Union's (ITU) IMT-2020 requirements, which entitled it to be referred to as 5G.   

\gls{3gpp} in Rel~15 designed 5G \gls{nr} to address a variety of usage scenarios requiring enhanced data rates, latency, coverage, and reliability. The key features of \gls{nr} include ultra-lean design, spectrum flexibility including operation in high frequency bands, interworking between high and low-frequency bands, and advanced antenna technologies. Rel~16 focused on \gls{urllc} and \gls{iiot}-related enhancements, \gls{nr} on unlicensed bands (\gls{nr}-U), and \gls{nr} \gls{v2x}. Recently, \gls{3gpp} completed Rel~17 which includes enhanced support for \gls{iiot}, proximity services, and network automation, as well as \gls{sl} enhancements for \gls{v2x} and public safety. It extends \gls{nr} operations to frequencies beyond 52~GHz, which is anticipated to lead to specifications in Rel~18. Figure~\ref{fig:Evolution} summarizes the evolution of radio access technologies from \gls{lte} to \gls{nr} and beyond.
\section{\gls{lte}-\gls{nr} Coexistence}
\label{sec:coexistence}
\begin{figure*}[t]
    \centering
    \includegraphics[width=\textwidth]{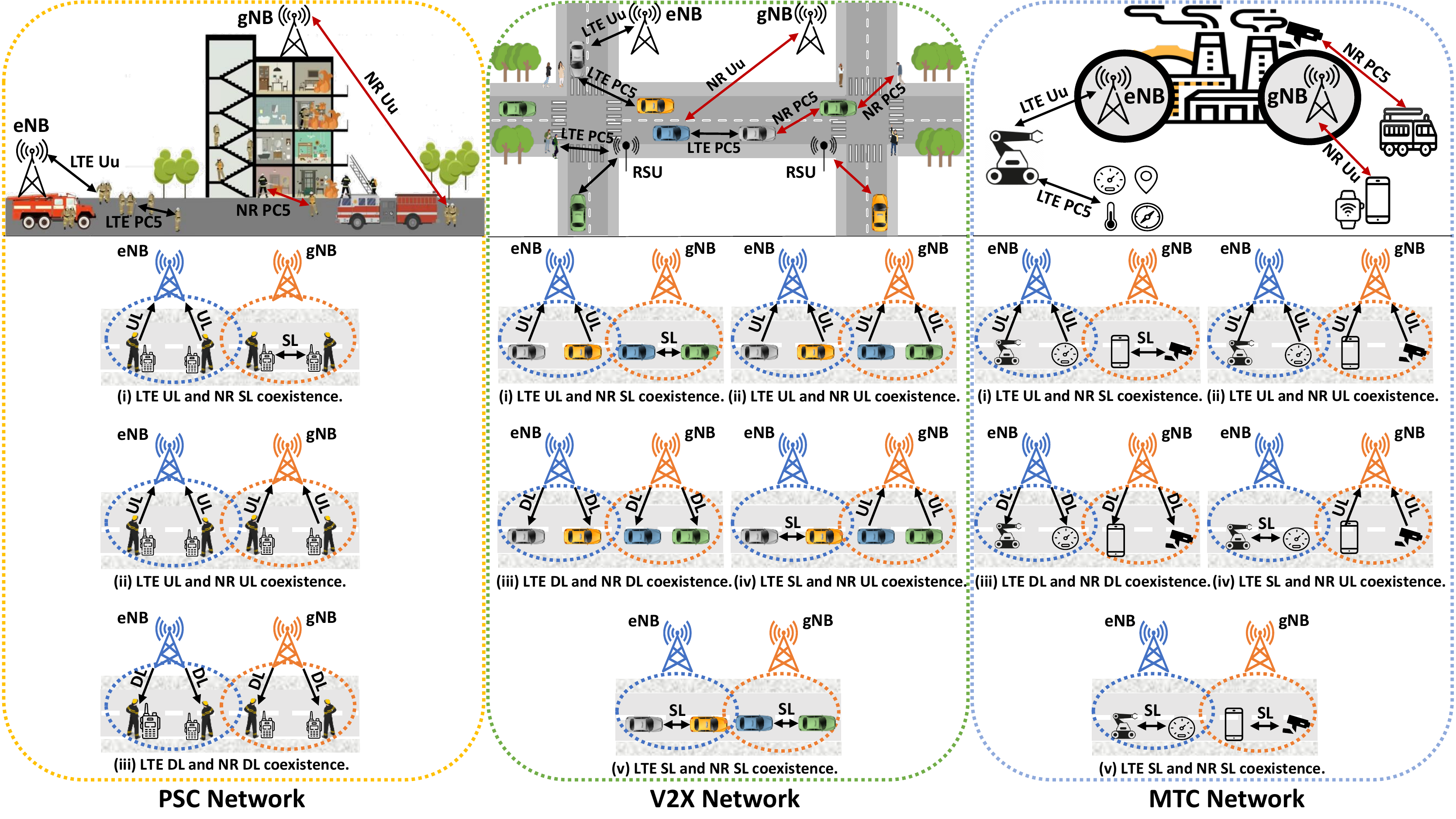}
    \caption{5G \gls{nr} applications and corresponding \gls{lte}-\gls{nr} coexistence scenarios. PC$5$ interface supports \gls{sl} communication and Uu interface supports \gls{ul} and \gls{dl}.}
    \label{fig:Coexistence}
\end{figure*}

5G \gls{nr} will operate in FR1 and FR2 bands, and coexist with legacy users, including \gls{lte} and radar systems.In this article, we consider only \gls{lte}-\gls{nr} coexistence, with a focus on different applications of \gls{embb}, \gls{mmtc}, and \gls{urllc}, i.e., \gls{psc}, \gls{mtc}, and \gls{v2x}, along with solutions proposed by \gls{3gpp}. Figure~\ref{fig:Coexistence} shows example \gls{psc}, \gls{mtc}, and \gls{v2x} networks and associated \gls{lte}-\gls{nr} coexistence scenarios.

\textbf{\gls{lte}-\gls{nr} Coexistence in \gls{psc} Networks:} \gls{psc} networks exist to protect people, maintain order, and facilitate recovery operations during emergencies, both natural and man-made. In addition to \gls{mtc} and \gls{urllc} services, the network relies on \gls{embb} services supported by \gls{nr}, which must coexist with \gls{lte}. To support coexistence, \gls{3gpp} introduced an \gls{lte}-compatible \gls{nr} numerology based on 15~kHz subcarrier spacing, which enables identical time/frequency resource grids for \gls{nr} and \gls{lte}. Flexible \gls{nr} scheduling, with a granularity at the subframe, slot, or symbol level, can also be employed to avoid collisions between \gls{nr} transmissions and \gls{lte} signals such as Cell Specific Reference Signal (CRS), signals/channels used for \gls{lte} initial access, and Channel State Information Reference Signal (CSI-RS). \gls{3gpp} also proposed resource reservation by competing systems to prevent collisions. 

The research community has studied \gls{lte}-\gls{nr} coexistence in \gls{embb} services in~\cite{ref:Coex_demmer_2018,ref:Coex_Xu_2021,ref:Coex_Alexandre_2019,ref:Coex_wan_2018,ref:Coex_Levanen_2018,ref:Coex_An_2020,ref:Coex_li_2021,ref:Coex_Wan_2019}. In~\cite{ref:Coex_demmer_2018}, the authors investigated the \gls{nr} numerology coexistence issue and evaluated the guard band size required to meet a target signal quality. In~\cite{ref:Coex_Xu_2021}, the authors investigated co-channel interference between non-colocated \gls{lte} and \gls{nr} systems and proposed a novel low-complexity CRS inteference mitigation algorithm for \gls{nr} \gls{ue}. In~\cite{ref:Coex_Alexandre_2019}, authors reported the experimental results corresponding to \gls{lte}-\gls{nr} coexistence in 700 MHz band and demonstrated peaceful \gls{dl} coexistence between the two technologies. In~\cite{ref:Coex_wan_2018}, the authors introduced a spectrum exploitation mechanism that balances the requirements of high transmission efficiency, large coverage area, and low latency. In~\cite{ref:Coex_Levanen_2018}, authors discussed \gls{lte}-\gls{nr} \gls{ul} coexistence and showed that from \gls{lte} system perspective a single resource block is sufficient as guard band to operate efficiently in the presence of \gls{nr} signal, whereas, \gls{nr} needs no guard band and is not affected by \gls{lte}. 

In~\cite{ref:Coex_An_2020}, authors explored the coexistence of \gls{lte} \gls{fdd} and \gls{nr} \gls{fdd} in 2.1~GHz band and showed that \gls{lte} \gls{fdd} \gls{dl} causes harmful interference to \gls{nr} \gls{fdd} \gls{dl}. In contrast to~\cite{ref:Coex_An_2020}, authors in~\cite{ref:Coex_li_2021} analyzed the coexistence of \gls{nr} \gls{fdd} and \gls{lte} \gls{tdd} in 1.8~GHz band and showed that these two technologies can coexist in adjacent bands. Lastly, in~\cite{ref:Coex_Wan_2019}, authors studied the coexistence of \gls{nr} \gls{ul}/\gls{dl} with \gls{lte} \gls{ul}/\gls{dl} and showed that \gls{ul} sharing helps achieve a balance between spectrum efficiency and low latency and coverage and channel bandwidth.

\textbf{\gls{lte}-\gls{nr} Coexistence in \gls{mtc} Network:}
\gls{3gpp} specified two low-power wide-area technologies, \gls{emtc} and \gls{nbiot}, to handle machine-centric use cases in 4G \gls{lte}. While \gls{emtc} is for mid-range machine-type applications and can support voice and video services, \gls{nbiot} supports low-cost devices and provides good coverage. Assessment of these technologies showed that each can satisfy \gls{nr} \gls{mtc} requirements; thus, \gls{3gpp} decided to reuse them to support \gls{nr} \gls{mtc} services within the \gls{nr} carrier. This is beneficial because \gls{emtc} and \gls{nbiot} devices are expected to remain in service even after \gls{lte} spectrum transitions to \gls{nr} and reusing the technologies will support legacy \gls{lte} devices. However, it will also lead to \gls{lte}-\gls{nr} coexistence issues in \gls{mtc} networks. 

To tackle this issue, \gls{3gpp} in TR (Technical Report)~37.823 investigated \gls{emtc} deployment within \gls{nr} carrier and proposed two solutions: resource reservation and \gls{dl}~subcarrier puncturing. The first allows reserving unused \gls{emtc} \gls{dl} and \gls{ul} resources for \gls{nr} signals at the subframe, slot or symbol level in the time domain and at the resource-block (RB) group level in the frequency domain. The second allows up to 2~subcarriers of \gls{emtc} \gls{ue} to be punctured to achieve RB alignment between \gls{emtc} and \gls{nr}.

In TR~37.824, \gls{3gpp}  investigated the coexistence issue when \gls{nbiot} operates: (a) in \gls{nr} in-band, (b) in \gls{nr} guard-band, and (c) independently of \gls{nr}. While no coexistence issue exists when \gls{nbiot} operates independently of \gls{nr}, \gls{3gpp} proposed resource allocation on non-anchor carriers to support coexistence when \gls{nbiot} operates in \gls{nr} in-band and guard-band. Resource reservation allows unused \gls{nbiot} \gls{dl} and \gls{ul} resources to accommodate \gls{nr} signals. The reservation is in time domain and can be at the subframe, slot, or symbol level.

The issue of coexistence between \gls{emtc}/\gls{nbiot} and \gls{nr} has also been studied in \cite{ref:Coex_Mozaffari_2019, ref:Coex_Ratasuk_2019, ref:Coex_Ratasuk_2020}. In~\cite{ref:Coex_Mozaffari_2019}, authors analyzed subcarrier grid and RB alignment problem and developed a framework for efficient placement of \gls{emtc}/\gls{nbiot} \gls{dl} carriers in \gls{nr} to avoid inter-carrier interference. In~\cite{ref:Coex_Ratasuk_2019} and~\cite{ref:Coex_Ratasuk_2020}, authors discussed the deployment of \gls{emtc}/\gls{nbiot} within the \gls{nr} carrier, the resulting coexistence issues, and mechanisms to alleviate them.

\textbf{\gls{lte}-\gls{nr} Coexistence in \gls{v2x} Network:}
\gls{3gpp} developed \gls{nr} \gls{v2x} to supplement \gls{lte} \gls{v2x} rather than supplant it, by providing support for advanced applications that the \gls{lte} system cannot carry. Since \gls{lte} \gls{v2x} has been standardized and is being deployed, it is likely that \gls{lte} \gls{v2x} and \gls{nr} \gls{v2x} will coexist. The coexistence can be within a single device, also known as in-device coexistence, or it can be between vehicles with \gls{lte} and \gls{nr} capabilities.    

To tackle the issue of \gls{lte}-\gls{nr} coexistence in \gls{v2x} networks, \gls{3gpp} in TR~37.985 proposed \gls{fdm} and \gls{tdm} solutions. With \gls{fdm}, a vehicle can simultaneously transmit over both \gls{lte} and \gls{nr}. However, the maximum transmit power must be shared between both technologies. To do so, \gls{3gpp} proposed dynamic and static power sharing. \gls{fdm} solutions can  be intra-band or inter-band. For intra-band operation, \gls{3gpp} proposed a static power allocation that is feasible for resolving conflicts due to simultaneous transmissions (Tx/Tx) from both technologies or when the transmission of one technology overlaps the reception of another (Tx/Rx), if the band separation is large enough. For inter-band operation, \gls{3gpp} proposed dynamic power sharing, which is feasible only if \gls{nr} and \gls{lte} transmissions fully overlap in time domain such that the total power across the transmissions is constant. 

In contrast to \gls{fdm}, if \gls{tdm} solutions are implemented, vehicles cannot simultaneously transmit over both technologies. \gls{3gpp} defined long term and short-term \gls{tdm} solutions. In long term solution, non-overlapping resource pools are pre-configured for \gls{nr} and \gls{lte} signals. In short term solution, provided the load for the vehicle from \gls{lte} and \gls{nr} side is at or below an acceptable level and assuming \gls{lte} employs semi-persistent scheduling, for each occurrence of Tx/Tx and Tx/Rx overlap, access technology prioritization is employed.

The issue of \gls{lte}-\gls{nr} coexistence in \gls{v2x} networks has also been addressed in~\cite{ref:V2X_Garcia_2021} and \cite{ref:V2X_Naik_2019}. While in~\cite{ref:V2X_Garcia_2021} authors focussed only on cellular technologies, i.e., \gls{lte}  and \gls{nr} \gls{v2x}, in~\cite{ref:V2X_Naik_2019}, authors discussed both cellular and Wi-Fi technologies along with their predecessors, i.e., IEEE 802.11bd with \gls{dsrc} and \gls{nr} with \gls{lte} \gls{v2x}. Authors provided an in-depth description of the technologies, 
described their key features, and discussed the issue of coexistence between them.
\section{Spectrum Sharing Scheme}
\label{sec:scheme}
Spectrum is the key enabler for \gls{nr} features such as enhanced data rates, ultra-low latency, and good coverage. However, to attain these features \gls{nr} has to efficiently share spectrum with \gls{lte}. In this section, we propose a spectrum sharing scheme for coexisting \gls{lte} and \gls{nr} networks. The proposed scheme uses a high-level model that makes few assumptions about the underlying technology and is therefore applicable to a general class of spectrum sharing scenarios beyond \gls{lte}-\gls{nr} coexistence. 

In general, let $N_G$ denote the set of coexisting networks. For \gls{lte}-\gls{nr} coexistence, $N_{G} = \{\ranA,\ranB\}$, where, $\ranA$ and $\ranB$ corresponds to an \gls{lte} and an \gls{nr} network, respectively. We assume that both networks use a shared pool of time-frequency resources and at regular intervals, the network operator measures the networks' respective demand for resources, so that each network's demand history forms a discrete-time random process. The intervals between measured demand levels can represent time slots in an \gls{lte} system, or $1$~ms subframes, or periods of several seconds, depending on the scenario being modeled.

Let $N_R$ denote the resource pool size that the network operator uses to allocate $N_A$ and $N_B$ resources to $\ranA$ and $\ranB$, respectively. The proposed scheme determines the optimal allocation that satisfies the networks' randomly varying demands by defining a metric to assess the performance of a particular allocation. 
The scheme can support static or dynamic (i.e., time-varying) allocations. 

When partitioning the pool, we constrain each network's allocation and the total allocation to lie in the interval $[0,N_r]$. We define a function $J$ that maps the ordered pair $(N_A,N_B)$ defined by the pool partition, to a scalar value that measures the performance of the network operator's partitioning of the pool. The objective is to find the partition that minimizes $J$, given the set of constraints. We choose $J$ based on the normalized Euclidean distance between $(x_A,x_B)$, an ordered pair containing statistics associated with the distributions of $\ranA$'s and $\ranB$'s respective demands, and $(N_A,N_B)$. We define the function as
\begin{align*}
\small
J= \gamma\left(\frac{N_A-x_A}{x_A}\right)^2 
+ (1-\gamma) \left(\frac{N_B-x_B}{x_B}\right)^2,
    \label{eqn:objective_function}
\end{align*}
where, $\gamma$ is a weighting factor that allows the network operator to prioritize one network over the other, if desired, and $(N_A-x_A)/x_A$ and $(N_B-x_B)/x_B$ are the fractional resource surpluses or deficits experienced by $\ranA$ and $\ranB$, respectively. While a surplus refers to over-provisioning or excess of resources, which could be used if the demand exceeds the estimated value, a deficit refers to under-provisioning or lack of resources and results in lost or buffered packets. Our goal is to minimize the surplus/deficit for both networks.

\begin{figure*}
     \centering
     \begin{subfigure}[b]{0.32\textwidth}
         \centering
         \includegraphics[width=\textwidth]{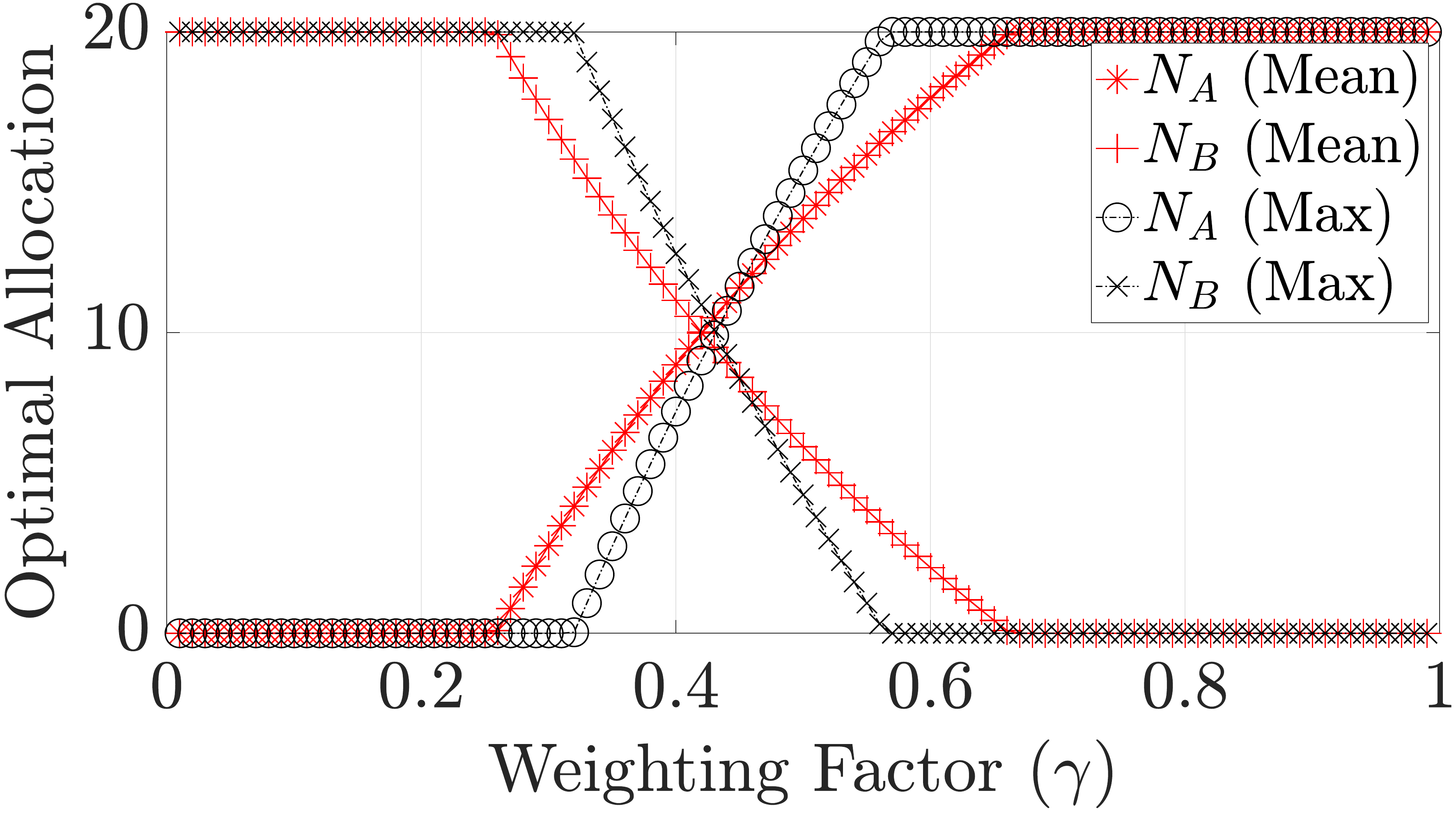}
         \caption{Optimal Allocation}
         \label{fig:Optimal_N_20}
     \end{subfigure}
     \hfill
     \begin{subfigure}[b]{0.32\textwidth}
         \centering
         \includegraphics[width=\textwidth]{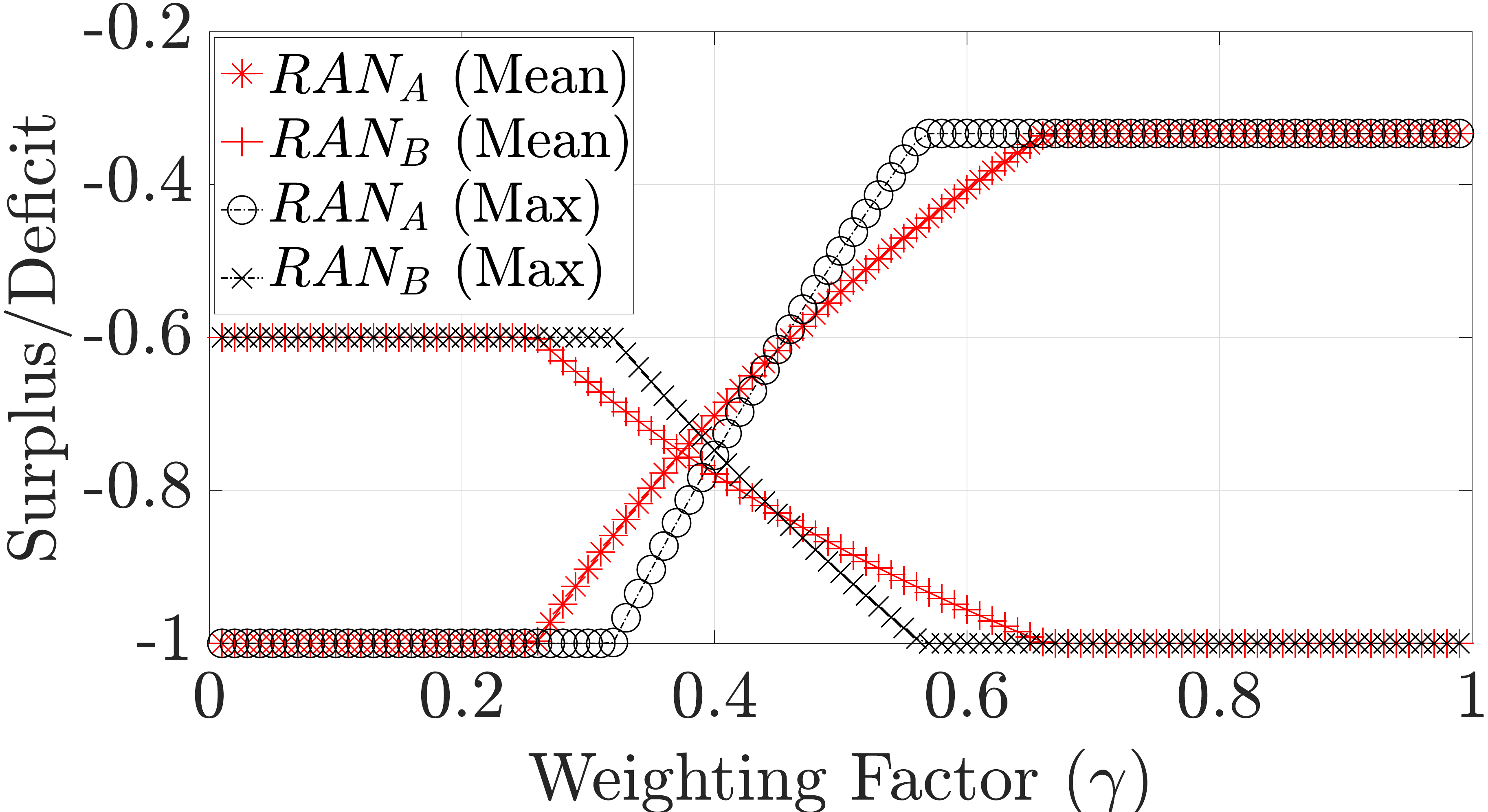}
         \caption{Surplus/Deficit}
         \label{fig:Surplus_Deficit_N_20}
     \end{subfigure}
     \hfill
     \begin{subfigure}[b]{0.32\textwidth}
         \centering
         \includegraphics[width=\textwidth]{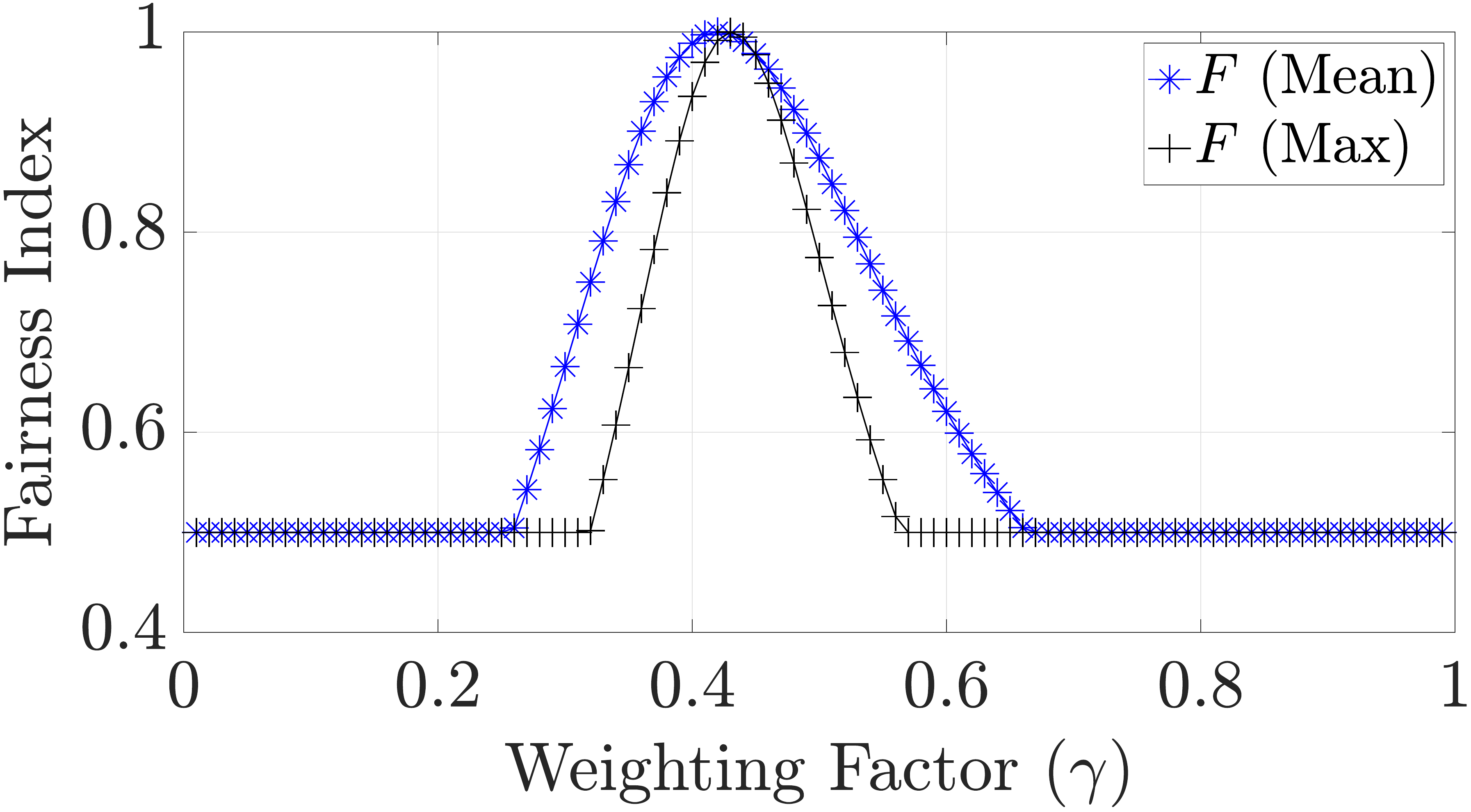}
         \caption{Jain's Fairness Index}
         \label{fig:Jain_Fairness_N_20}
     \end{subfigure}
    \caption{Results for maximum demand versus mean demand optimization, for resource pool size, $N_R = 20$.}
        \label{fig:results}
\end{figure*}

The defined constrained function can be employed by the network operator in multiple ways. For instance, if the networks are capable of providing accurate predictions of their demand in the next time period, the operator can use the constrained function periodically and adjust the resource pool partition accordingly. Alternatively, the operator can use the constrained function to allot a static partition of the resource pool at network setup. However, the latter assumes that the demand processes are stationary which may not be applicable to scenarios such as one where network sizes vary. 

In this article, we discuss static resource allocation. Specifically, we consider two scenarios: one that uses the maximum demand from each network as the input statistic to the constrained function and another that uses the mean demand. We then examine the effect of the pool size relative to the networks' demand on the outcome of the optimization problem for both scenarios. Lastly, we compute the optimal resource pool partition $(N_A^*,N_B^*)$, the resulting fractional surplus/deficit for each network, and Jain's Fairness Index for the partition scheme, which is $F = (N_A^*+N_B^*)^2/2((N_A^*)^2+(N_B^*)^2)$.

We use Autoregressive Moving-Average (ARMA) random processes to model the resource demands from the networks. The ARMA processes are examples and are not based on data from physical devices. Since the demand process statistics may not be known {\it a priori}, we generated $1000$ realizations of the ARMA process and from these observations computed $95$~\% confidence intervals for the mean, variance, and maxima.

For our analysis, we considered three resource pool sizes: $N_R = \{20,60,100\}$. For each pool size, we varied the parameter $\gamma$ from $0$ to $1$. For each value of $\gamma$, we used the ARMA process statistics from $\ranA$ and $\ranB$ to obtain the pairs of optimal partition sizes $N_A^*$ and $N_B^*$ associated with both the upper and lower bounds of the $95$~\% confidence intervals of the process' means, variances, and maxima. With the optimal allocations based on process statistics (means and variances) and process maxima that are associated with the upper and lower confidence interval bounds in hand, we computed the corresponding ranges of the expected fractional surplus or deficit experienced by each process, as well as the corresponding Jain's fairness index, for each value of $\gamma$ and each resource pool size, $N_R$. 

The measured $95$~\% confidence intervals for the $\ranA$ and $\ranB$ demand processes are as follows. For the process means, $\mu_A = [29.96, 30.03]$ and $\mu_B = [49.60, 50.36]$, for the process variances, $\sigma_A^2 = [20.30, 20.62]$ and $\sigma_B^2 = [28.42, 31.06]$, and for the maxima, $P_A = [43, 43]$ and $P_B = [62, 63]$. Thus, the expected total demand is approximately $80$~resources while the maximum total demand is approximately $105$~resources. In the interest of space, we only shows the results for $N
_r = 20$ in Figure~\ref{fig:results}. Also, since the $95$~\% confidence interval values for the statistics are very close, we show results corresponding only to the lower interval values.

When $N_r=20$, the pool size is much smaller than the estimated demand, and starvation is an issue for both networks. Total starvation (i.e., the network receives no resources) corresponds to a fractional deficit of $-1$. From Figure~\ref{fig:Surplus_Deficit_N_20}, the range of values for $\gamma$ over which both networks experience total starvation is smaller when we use the statistics. Also, Figure~\ref{fig:Jain_Fairness_N_20} shows that using the statistics produces a fairer allocation over a wider range of $\gamma$.

When $N_r=60$, while the pool size is more than the mean demand of both the networks, it is close to the peak demand of $\ranB$ and more than that of $\ranA$. Using the statistics-based optimization allows us to achieve a perfectly fair allocation when $\gamma \approx 0.55$, although the maxima-based optimization is fairer when one network has much higher priority than the other, i.e. when $\gamma$ is very close to 0 or 1, and results in an optimal allocation that leads to surpluses or over-provisioning of resources. In this scenario, the scheduler can use the maxima to partition the pool when one network has higher priority than the other.

Lastly, when $N_r=100$, the pool is sufficient to meet the mean demands, and is large enough to support the maximum total demand. In this case, using the maxima-based optimization performs better for both networks, and results in an allocation that leads to surplus, i.e., over-provisioning, of resources. Thus, as the resource pool size gets closer to the maximum total demand, using the process mean and variance results in an allocation equal to the mean demand and results in a deficit, while using the maxima results in over-provisioning. 

In future, we will consider other stationary and non-stationary demand models and develop a learning-based resource allocation scheme that employs the proposed optimization strategy. 
\section{Conclusions}
In this article, we discussed \gls{lte}-\gls{nr} coexistence for \gls{embb}, \gls{mmtc}, and \gls{urllc} use cases in \gls{psc}, \gls{mtc}, and \gls{v2x} networks. We examined how \gls{3gpp} is developing solutions for performance improvement when \gls{nr} and \gls{lte} devices coexist and listed some approaches. Lastly, we introduced a novel and generic spectrum sharing scheme that is applicable to a variety of use cases and which allows for static resource allocation. 
\section*{Acknowledgement}
The authors thank the reviewers and Chen Shen (Georgetown University and NIST, presently Google) for their valuable comments.
\section*{Biography}
\label{sec:bio}
\small
\textbf{Sneihil Gopal} received her Ph.D. degree in Electronics and Communications Engineering in $2021$ from IIIT-Delhi, India. She is currently working as a Postdoctoral Researcher at the Department of Physics, Georgetown University, and as an International Associate Researcher in the Wireless Networks Division, NIST. Her research interests include dynamic spectrum sharing, wireless network optimization, applications of game theory in wireless networks, and machine learning. 

\textbf{David Griffith} is with the Wireless Networks Division at NIST. He received the Ph.D. degree in electrical engineering from the University of Delaware in 1998. He has authored or co-authored nearly 100 publications, including two book chapters, on non-linear signal processing, satellite communications, optical communications, smart grid communications, public safety communications, and machine learning for communications in the Industrial Internet of Things (IIoT). His work includes modeling wireless communication systems and machine learning systems. 

\textbf{Richard Rouil} received his Ph.D. degree in computer science in 2009 from Telecom Bretagne, France, that focused on mobility in heterogeneous networks. He is currently the Division Chief of the Wireless Networks Division at NIST. His research focuses on the performance evaluation of wireless technologies, such as \gls{lte} and \gls{nr} to support the development, analysis, and deployment of networks used by public safety. His main interests include protocol modeling and simulation of communication networks.

\textbf{Chunmei Liu} received her Ph.D. degree in Computer Systems from the Massachusetts Institute of Technology (MIT) in 2005. She has been working in the areas of wireless communications and networking since then. She is currently a researcher with the Wireless Networks Division, NIST. Her research interests include public safety communications in cellular networks and machine learning.
\begin{spacing}{}
\bibliographystyle{IEEEtran}
\bibliography{references}
\end{spacing}
\end{document}